\documentclass[aps,prd,twocolumn,showpacs,groupedaddress,floatfix]{revtex4-1}

\usepackage{graphicx}
\usepackage{array}
\usepackage{amsmath}
\usepackage{amssymb}
\usepackage{amsfonts}

\DeclareGraphicsExtensions{.eps, .eps, .jpg, .png}

\newcommand{\bea}{\begin{eqnarray}}
\newcommand{\eea}{\end{eqnarray}}
\newcommand{\be}{\begin{equation}}
\newcommand{\ee}{\end{equation}}
\newcommand{\bi}{\begin{itemize}}
\newcommand{\ei}{\end{itemize}}

\newcommand{\ie}{{\it i.e.}}
\newcommand{\eg}{{\it e.g.}}
\newcommand{\nn}{\nonumber}

\newcommand{\pt}{\partial}

\def\d{{\rm d}}

\def\cH{{\cal H}}

\def \fnl{f_{\rm N\hspace{-.5pt}L}}

\begin{document}

\title{Sudden variations in the speed of sound during inflation:\\  features in the power spectrum and bispectrum}

\author{Minjoon Park}
\email[]{minjoonp@physics.umass.edu}
\affiliation{Department of Physics, University of Massachusetts, Amherst, MA 01003}

\author{Lorenzo Sorbo}
\email[]{sorbo@physics.umass.edu}
\affiliation{Department of Physics, University of Massachusetts, Amherst, MA 01003}

\date{\today}
\begin{abstract}

We employ the formalism of the effective field theory of inflation to study the effects of a sudden change in the speed of sound of the inflationary perturbations. Such an event generates a feature with high frequency oscillations both in the two- and in the three-point functions of the curvature fluctuations. We study, at first order in the magnitude of the change of the speed of sound, the dependence of the power spectrum and of the bispectrum on the duration of the change. In the limit of a very short duration, the oscillations in the power spectrum persist up to very large momenta and the amplitude of the feature in the bispectrum diverges while its location moves to increasing momenta.

\end{abstract}

\maketitle

%%%%%%%%%%%%%%%%%%%%%%%%
\section{Introduction}%%
%%%%%%%%%%%%%%%%%%%%%%%%

Inflation, in its simplest realizations, is a period of rather uneventful quasi-de Sitter expansion of the Universe. As a consequence, it typically predicts a quasi-scale invariant, featureless spectrum of scalar and tensor metric perturbations. It is however possible that the inflating Universe did experience some traumatic events that left a permanent mark on the perturbations we currently observe in the Cosmic Microwave Background radiation.

Various examples of such traumatic events have been discussed in the past. The simplest possibility is that they originate from a feature,  possibly associated to a phase transition~\cite{Adams:1997de}, in the shape of the inflationary potential~\cite{Adams:2001vc,Kaloper:2003nv,Arroja:2011yu,Martin:2011sn,Adshead:2011jq}. Another possibility is that the slowly rolling inflaton induces an explosive production of quanta of some other degree of freedom~\cite{Chung:1999ve,Romano:2008rr,Barnaby:2009mc} that affect the metric perturbations. In general these phenomena lead to high frequency oscillations in the primordial power spectrum about the scales that left the horizon at the time of the event. Several groups -- see for instance~\cite{Martin:2003sg,Benetti:2011rp,Meerburg:2011gd} --  have found hints of the presence of such oscillations in the CMB data.

In the present paper we study a different kind of sudden event: a change in the speed of sound of the inflaton perturbations. The speed of sound $c_s$ of the inflaton can be different from the speed of light in models with nonstandard kinetic terms~\cite{ArmendarizPicon:1999rj}, and in particular in DBI inflation~\cite{Alishahiha:2004eh}. It is therefore natural to ask what observable effects can occur when the speed of sound suddenly changed during inflation. In the past, this situation has been considered in~\cite{Bean:2008na} and in~\cite{Nakashima:2010sa}, which studied the effect of a jump in $c_s$ on the two-point function of the inflaton. Reference~\cite{Bean:2008na} did also estimate the effect of such a jump on the bispectrum.

In our work we provide a detailed study of how a rapid but not instantaneous change in $c_s$ affects both the power spectrum and the bispectrum of the curvature perturbations. The Lagrangian provided by the effective field theory of the inflaton fluctuations~\cite{Cheung:2007st} serves our purpose well, in that it allows us to determine the effect of the jump in the speed of sound in a model-independent way. Differently from what happens in most of the literature, we study the system of~\cite{Cheung:2007st} without assuming that the background parameters evolve adiabatically. 

Since we work in the context of~\cite{Cheung:2007st}, we are not interested in the details of the full theory producing the sudden change in $c_s$, and will not worry about the issue of its technical naturalness. Let us just note that reference~\cite{Bean:2008na} discusses an explicit example where brane inflation leads to a jump in $c_s$. Reference~\cite{Nakashima:2010sa} contains a couple of additional examples. And here we mention one more possibility: In general, the speed of sound in DBI inflation~\cite{Alishahiha:2004eh} depends on the expectation value of several fields~\cite{Langlois:2008wt}. The rolling inflaton can lead,  \`a la~\cite{Chung:1999ve}, to the nonperturbative production of matter in the sector determining the value of $c_s$. If this occurs, then the speed of sound will momentarily change its value before the nonperturbatively produced particles dilute away.

Let us now summarize our main results. The background dynamics producing a change in the speed of sound will also induce, in general, a change in the slow roll parameters $\epsilon$ and $\eta$. In order to disentangle the different effects, however, we will consider a scenario where $\epsilon$ and $\eta$ are kept constant and only $c_s$ changes. Now, as long as both the speed of sound and the slow roll parameters are constant, the amplitude of the power spectrum is inversely proportional to $c_s$ and to $\epsilon$. The spectral index, on the other hand, depends on $\epsilon$ and $\eta$, but not on $c_s$. As a consequence, a sharp jump in $c_s$ will induce a jump in the amplitude of the power spectrum without affecting its tilt. Our analytical and numerical work will confirm this expectation. This property might be used to discriminate models with a sudden change in the inflationary potential from those with a sudden change in $c_s$. One might even be tempted to use a sudden change in the speed of sound to explain the small amplitude of the observed CMB fluctuations at the largest multipoles.

Our system depends on two parameters -- the change $\Delta c_s$ in the speed of sound and the duration $\Delta \tau$ of such a change. As long as $\Delta \tau$ is not vanishing, the two-point function shows an oscillating pattern that extends over a finite range of momenta. The duration of the oscillations becomes infinite in the limit $\Delta \tau\to 0$, in agreement with the results of~\cite{Bean:2008na,Nakashima:2010sa}, reflecting a UV artifact that originates from the discontinuous behavior of $c_s$. The three-point function or $\fnl$, when evaluated in the equilateral configuration, also oscillates, with a maximum amplitude scaling as $\Delta c_s\,\tau_0^2/\Delta\tau^2$ at momentum $\sim\Delta\tau^{-1}$, where $\tau_0$ is the (conformal) time when the change in $c_s$ occurs (the amplitude of the bispectrum in the squeezed limit scales as $\Delta c_s\,|\tau_0|/\Delta\tau$, and agrees with the consistency relation~\cite{Creminelli:2004yq}). Therefore a jump that is rapid enough may cause large nongaussianity at large momenta. The behavior of the bispectrum turns out to be similar to that produced by a sudden change in the inflaton potential recently studied in~\cite{Adshead:2011jq}, and future surveys might be able to detect it if the feature appears at sufficiently large scales.

The plan of the paper is the following. In section~\ref{gauge} we apply the gauge invariant formulation of cosmological perturbations to the setup of \cite{Cheung:2007st}, and derive a quadratic action for the gauge invariant variable $v$ describing the scalar perturbation. Then we model the jump in $c_s$ and solve the equation of motion for $v$, to obtain the expression of the power spectrum. In section~\ref{bispectra} we take the decoupling limit, study its regime of validity and calculate, in this limit, the equilateral and the squeezed bispectra for our system. Finally, in section~\ref{disc} we summarize our results and discuss future prospects.

%%%%%%%%%%%%%%%%%%%%%%%%%%%%%%%%%%%%%%%%%%%%%%%%%%%%%%%%%%%%%%%%%%%%%%
\section{Gauge invariant formulation at quadratic order}\label{gauge}%
%%%%%%%%%%%%%%%%%%%%%%%%%%%%%%%%%%%%%%%%%%%%%%%%%%%%%%%%%%%%%%%%%%%%%%

%%%
\subsection{Action}
%%%

We employ the effective field theory of inflation proposed by~\cite{Cheung:2007st}. Since we are interested only in changes in the speed of sound, we neglect the terms in $M_3$, $\bar{M}_2$ and $\bar{M}_3$, etc. of \cite{Cheung:2007st} and start with the following action
\begin{align}\label{eqn:ggact}
S = \int  \d^4 x & \sqrt{- g}\, \Big[ \frac{M_p^2}{2} R - c(t+\sigma)\tilde g^{00} \\
&- \Lambda(t+\sigma) + \frac{1}{2!}M_2(t+\sigma)^4(1+\tilde g^{00})^2\,\Big] \,,\nonumber
\end{align}
where $\sigma$ is the Goldstone field and 
\bea
\tilde g^{00} &=& \frac{\pt (t+\sigma)}{\pt x^\alpha} \frac{\pt (t+\sigma)}{\pt x^\beta} g^{\alpha\beta} \,.
\eea
Using the ADM formalism, we write the metric as $\d s^2 = -N^2\d t^2+\gamma_{ij}(\d x^i+N^i\d t)(\d x^j+N^j\d t)$, and expand at first order $N=1+\phi$, $N_i=\pt_i B$ and $\gamma_{ij}=a^2\big\{(1-2\,\psi)\delta_{ij}+2\pt_i\pt_j E\big\}$. Then the perturbative expansion of the action~(\ref{eqn:ggact}) is straightforward. At first order in perturbations we obtain the  following background equations
\begin{align}
&c+\Lambda-3M_p^2H^2=0\,,\nonumber\\
&c-\Lambda+M_p^2(3H^2+2\dot H)=0\,,
\end{align}
with an overdot implying a differentiation with respect to the coordinate time $t$ and $H=\dot a/a$.

Before we get to the second order, we write $\phi$ and $B$ in terms of the other field variables by perturbatively solving the lapse and shift constraints $\delta S/\delta N=0$ and $\delta S/\delta N^i=0$:
\begin{align}\label{eqn:subst}
\phi&=\epsilon\,H\,\sigma-\frac{\dot\psi}{H}\,,\\
\frac{\Delta B}{a^2}&=\frac{\Delta\psi}{a^2H}-\frac{\epsilon}{c_s^2}\,\dot\psi+\Delta\dot E+\frac{\epsilon^2}{c_s^2}\,H^2\,\sigma-\frac{\epsilon}{c_s^2}\,H\,\dot\sigma\,, \nonumber
\end{align}
where we have defined
\begin{align}
\epsilon\equiv-\frac{\dot H}{H^2}\,,\quad \mu\equiv\frac{2\,M_2^4}{M_p^2\,H^2}\,,
\end{align}
and where the speed of sound is given by 
\begin{equation}
c_s^2\equiv\frac{\epsilon}{\epsilon+\mu}\,.
\end{equation}
With (\ref{eqn:subst}) plugged in, the quadratic action depends only on $\psi$, $E$ and $\sigma$. The terms in $E$, however, can be dropped because they are total derivatives. Finally, we introduce the gauge invariant variable 
\be\label{eqn:giv}
v\equiv z\,(\psi+H\sigma)\,,
\ee
with
\begin{align}
z^2=2\,M_p^2\,a^2\,\left(\epsilon+\mu\right)\,,
\end{align}
so that the quadratic action, after dropping more total derivative terms, reads 
\be\label{eqn:vact}
S^{(2)}=\frac{1}{2}\int\d\tau\,\d^3x\Big\{(v')^2-c_s^2(\pt_iv)^2+\frac{z''}{z}v^2\Big\}\,,
\ee
where $\tau$ is the conformal time and ${}'\equiv\frac{\d}{\d\tau}$. The resulting mode equation for the Fourier transform $\hat{v}$ of $v$ is
\be\label{eqn:modeeq}
\hat v''+\Big(c_s^2\,k^2-\frac{z''}{z}\Big)\hat v=0\,.
\ee

%%%
\subsection{Modeling the non-adiabatic change of $c_s$}
%%%

We are interested in the effects of a non-adiabatic change in $c_s$ originating from that of $\mu$, {\em i.e.}, of $M_2$. We will consider a small jump in $\mu$ during a short time about $\tau=\tau_0$. Since the overall normalization of the power spectrum is inversely proportional to $c_s$, data allow only for a small jump $|\Delta c_s|\ll 1$ in the speed of sound. For simplicity we will assume that $c_s=1$ before the transition. Since for $c_s$ close to unity ({\em i.e.}, $\mu\ll \epsilon$) one has $c_s\simeq 1-\frac{\mu}{2\,\epsilon}$, we parametrize
\be\label{eqn:regmu}
\frac{\mu(\tau)}{\epsilon} = \left\{ \begin{array} {ll} 
0 \,, \quad  &\tau<\tau_0\,, \\ \\
-2\,\Delta c_s\cdot f(\tau)\,, \quad & \tau_0<\tau<\tau_0+\Delta\tau \,, \\ \\
-2\,\Delta c_s\,, \quad & \tau>\tau_0+\Delta\tau \,,
\end{array} \right.
\ee
with $f(\tau_0)=0$ and $f(\tau_0+\Delta\tau)=1$. Here the magnitude of the jump, $\Delta c_s$, has to be small in order to match observations, but for a sufficiently short $\Delta\tau$ the rate of change can be large enough to open the possibility of an observable signature. For the sake of simplicity, one may be tempted to choose $f$ to be linear in $\tau$, or even a step function. However, a discontinuity in $c_s$ or its derivatives can generate UV artifacts that contaminate the relevant information. For example, if $\mu$ jumps as a step function, then the feature coming from the sudden transition at $\tau=\tau_0$ persists in all scales $k>|\tau_0|^{-1}$ already at the two-point function level~\cite{Bean:2008na,Nakashima:2010sa}. With $f$ linear in $\tau$, such that $\mu$ is continuous but $\mu'$ is not, the oscillations in the power spectrum die off at large $k$, but the bispectrum is linearly divergent. It turns out that in order to obtain a vanishing bispectrum in the UV we must choose $\mu$ to be a function of class ${\mathcal C}^2$. 

%%%
\subsection{Mode function}\label{mf}
%%%

In order to solve eq.~(\ref{eqn:modeeq}) we employ the iterative method of \cite{Stewart:2001cd}, to compute the mode function as a power series in the small parameter $\Delta c_s$. We will denote by subscripts ${}_1$, ${}_2$ and ${}_3$ the time intervals of before, during and after the period of non-adiabatic change in $c_s$, respectively. 

Before the transition ($\tau<\tau_0$), eq.~(\ref{eqn:modeeq}) becomes
\be\label{eqn:region1}
\hat v_1''+\Big(k^2-\frac{2+3\,(2\,\epsilon-\eta)}{\tau^2}\Big)\,\hat v_1=0\,,
\ee
where $\eta\equiv-\frac{\ddot{H}}{2\,H\,\dot{H}}=\epsilon-\frac{\epsilon'}{2\,\cH\,\epsilon}$ is the second slow-roll parameter. Assuming $\epsilon'$ to be negligible, so that $\eta\simeq\epsilon$, the solution of eq.~(\ref{eqn:region1}) with the conventional boundary condition, $\hat v_1(\tau\to-\infty)=e^{ik\tau}/\sqrt{2k}$, is
\be\label{eqn:v1}
\hat v_1 = -\frac{\sqrt{-\pi\tau}}{2}H^{(1)}_{\frac{3}{2}+\epsilon}(-k\tau)\,.
\ee
Next, we introduce the dimensionless variables $x\equiv-k\,\tau$ (so that $0<x<+\infty$, $x_0\equiv -k\,\tau_0$ and $\Delta x\equiv k\,\Delta\tau$) and $s\equiv\sqrt{2\,k}\,\hat v$. In terms of these variables eq.~(\ref{eqn:v1}) reads 
\be
s_1(x) = -\sqrt{\frac{\pi x}{2}}H^{(1)}_{\frac{3}{2}+\epsilon}(x)\,.
\ee

During the transition($\tau_0<\tau<\tau_0+\Delta\tau$), we write the mode equation as
\begin{equation}\label{eqn:region2}
\frac{\d^2s_2}{\d x^2}+\Big(1-\frac{2+3\epsilon}{x^2}\Big)\,s_2(x)=2\,\Delta c_s\, p(x)\,s_2(x) + \dots \,,
\end{equation}\\
where $\dots$ denotes, from here on, any ${\cal O}(\epsilon^2,\Delta c_s^2,\epsilon\,\Delta c_s)$ correction and 
\bea
p(x)&=&-f(x)+\frac{1}{x}\frac{\d f}{\d x}-\frac{1}{2}\,\frac{\d^2f}{\d x^2} \,.
\eea
Requiring that $s_2$ satisfies the boundary conditions $s_2(x_0)=s_1(x_0)$ and $s_2'(x_0)=s_1'(x_0)$ fixes the homogeneous solution of (\ref{eqn:region2}) to be just $s_1(x)$. Using the standard Green's function technique, we obtain
\begin{align}\label{eqn:region2sol}
s_2(x) = s_1(x)+i&\,\Delta c_s\int_x^{x_0}\d y \, p(y)\,s_1(y)\\
&\times\big\{s_1^*(y)\,s_1(x)-s_1^*(x)\,s_1(y)\big\} + \dots \,.\nonumber
\end{align}

For $\tau>\tau_0+\Delta\tau$, the mode equation reads
\be\label{eqn:region3}
\frac{\d^2s_3}{\d x^2}+\Big(\frac{1}{1-2\,\Delta c_s}-\frac{2+3\epsilon}{x^2}\Big)\,s_3(x)=0\,,
\ee
whose solution is 
\bea
s_3 &=& -\sqrt{\frac{\pi x}{2}}\Big\{\alpha\,H^{(1)}_{\frac{3}{2}+\epsilon}\Big(\frac{x}{\sqrt{1-2\,\Delta c_s}}\Big) \nn\\
&&\hspace{40pt}+ \beta\,H^{(2)}_{\frac{3}{2}+\epsilon}\Big(\frac{x}{\sqrt{1-2\,\Delta c_s}}\Big)\Big\}\,.
\eea
The coefficients $\alpha$ and $\beta$ are determined by matching $s_3$ and $s_2$ smoothly at $x=x_0-\Delta x$: $s_3(x_0-\Delta x)=s_2(x_0-\Delta x)$ and $s_3'(x_0-\Delta x)=s_2'(x_0-\Delta x)$.

%%%
\subsection{Power spectrum}\label{ps}
%%%

To show concrete results, we choose our $f$ to be the simplest polynomial in $\tau$ that gives a ${\cal {C}}^2$ interpolation between $f(\tau_0)=0$ and $f(\tau_0+\Delta\tau)=1$:
\be\label{eqn:connect}
f(\tau)=\frac{(\tau-\tau_0)^3\big\{3(2\tau-2\tau_0-3\Delta\tau)(\tau-\tau_0-\Delta\tau)+\Delta\tau^2\big\}}{\Delta\tau^5}\,.
\ee
Omitting the gruesome details of the intermediate steps, we eventually obtain
\begin{align}
&\alpha=1-\frac{i}{2}\,\Delta c_s\,(2\,x_0-\Delta x) + \dots \,,\\
&\beta=\frac{45}{2}\,e^{i\,(2\,x_0-\Delta x)}\,\frac{\Delta c_s}{\Delta x^4}\,\big\{\cos \Delta x-(1-\frac{\Delta x^2}{3})\,\frac{\sin \Delta x}{\Delta x}\,\big\} +\dots\,,\nonumber
\end{align}
giving the power spectrum
%%%%
\begin{widetext}
\bea\label{eqn:sd}
{\cal P}(k)&=&\frac{k^3}{2\pi^2}\lim_{x\to 0}\frac{\left|s_3(x)\right|^2}{2\,k\,z(x)^2}=\frac{H^2}{8\pi^2M_p^2\,\epsilon}\Big\{1-\Delta c_s+2\epsilon\,\left(2-\gamma_E-\log 2k\right) +\dots\Big\}\,|\alpha-\beta|^2 \nn\\
&=&\frac{H^2}{8\pi^2M_p^2\epsilon}\Big\{1+2\epsilon\,\left(2-\gamma_E-\log 2k\right) \\
&&\hspace{45pt}-\Delta c_s\,\left[1+45\,\cos k(2\tau_0+\Delta\tau)\,\left(\frac{\sin k\Delta\tau}{3\,k^3\Delta\tau^3}+\frac{\cos k\Delta\tau}{k^4\Delta\tau^4}-\frac{\sin k\Delta\tau}{k^5\Delta\tau^5}\right)\right] +\dots\Big\} \,,\nn
\eea
\end{widetext}
%%%%
where $\gamma_E$ is the Euler-Mascheroni constant. The asymptotic behavior of the two-point function
\be
{\cal P}(k) = \left\{ \begin{array} {l} 
\frac{H^2}{8\pi^2M_p^2\epsilon}\left[1+2\epsilon\,\left(2-\gamma_E-\log 2k\right)\right] \,,  \\
\qquad \qquad\qquad \qquad\qquad\qquad|k\tau_0|\ll 1\,, \\ \\
\frac{H^2}{8\pi^2M_p^2\epsilon}\left[1-\Delta c_s+2\epsilon\,\left(2-\gamma_E-\log 2k\right) \right] \,, \\\qquad \qquad\qquad \qquad\qquad\qquad |k\tau_0|\gg 1 \,,
\end{array} \right.
\ee
agrees, as expected, with the spectral densities for $c_s=1$ and $c_s=1+\Delta c_s+{\cal O}(\Delta c_s^2)$ without any sudden transition. 

%%%%%%%%%%%%%%%
%%%%%%%%%%%%%%%
\begin{figure}
  \begin{center}
    \resizebox{.5\textwidth}{!}{\includegraphics{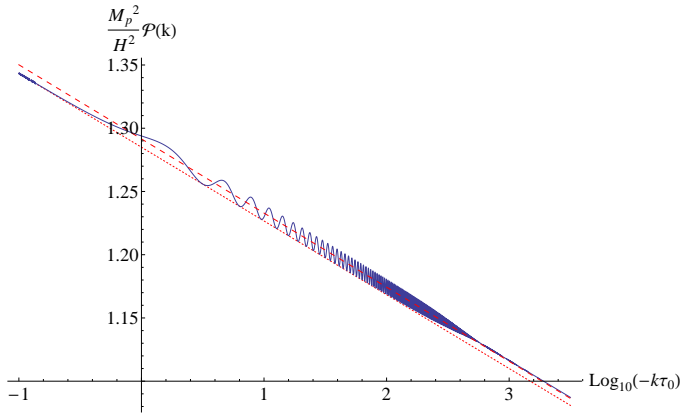}}
  \end{center}
  \caption{Spectral density for $\epsilon=.01$, $\Delta c_s=-.005$ and $\Delta\tau=-.01\,\tau_0$. Solid blue line: eq.~(\ref{eqn:sd}), dotted red: IR asymptotics, dashed red: UV asymptotics.}
  \label{sd}
\end{figure}
%%%%%%%%%%%%%%%
%%%%%%%%%%%%%%%

The feature generated by the non-adiabatic change in $c_s$ shows up in the term proportional to $\Delta c_s$ in (\ref{eqn:sd}): it vanishes for small $k$, as can be seen from the IR asymptotics. Oscillatory behavior of the amplitude of $\Delta c_s$ appears once we deviate from the IR regime, \ie, $-k\,\tau_0\gtrsim1$, but it dies off at $k\sim 1/\Delta\tau$ due to the negative powers of $k\,\Delta\tau$ in the oscillating terms. A plot of ${\cal P}(k)$ for $\epsilon=.01$, $\Delta c_s=-.005$ and $\Delta\tau=-.01\,\tau_0$ is shown in Fig.\ref{sd}.

%%%%%%%%%%%%%%%%%%%%%%%%%%%%%%%%%%%%%%%%%%%%%%%%%%%%%%%%%%
\section{Decoupling limit and bispectra}\label{bispectra}%
%%%%%%%%%%%%%%%%%%%%%%%%%%%%%%%%%%%%%%%%%%%%%%%%%%%%%%%%%%

In the previous section we have derived the two-point function of the perturbations by computing the full gauge invariant quadratic action with a varying speed of sound. In order to compute the three-point function of the perturbations we need to move on to the cubic Lagrangian for this system. We will use here the fact that a sufficiently rapid change in the speed of sound will  mostly affect the modes that are deep inside the horizon at $\tau\simeq \tau_0$. For the analysis of those modes it is sufficient to study the action~(\ref{eqn:ggact}) in the decoupling limit.

Let us first check the regime of validity of the quadratic Lagrangian in the decoupling limit. In this limit, the quadratic part of the Goldstone sector of (\ref{eqn:ggact}) is
\begin{equation}\label{eqn:sigact}
S_\sigma^{(2)}=  \int\d\tau\,\d^3x\,\frac{M_p^2\,\cH^2\,\epsilon}{c_s^2}\,\Big[\sigma'{}^2-c_s^2\,(\pt_i\sigma)^2-3\,\cH^2\,\epsilon\,\sigma^2\Big] \,,
\end{equation}
with $\cH\equiv a'/a$. Upon canonical normalization of $\sigma$ by $w=\frac{\cH z}{a}\sigma$, we get
\begin{align}\label{eqn:wact}
S_\sigma^{(2)} = &\frac{1}{2}\int\d\tau\,\d^3x\,\Bigg[w'{}^2-c_s^2\,(\pt_iw)^2+\frac{z''}{z}\,w^2 \nonumber\\
&-\left(2\,\cH\,\epsilon\,\frac{z'}{z}+\epsilon\,\cH^2\left(3\,c_s^2+1-2\,\eta\right)\right)w^2\Bigg]\,,
\end{align}
where the second line contains terms that should be neglected in the decoupling limit $k\gg \sqrt{\epsilon}\,\cH$~\cite{Cheung:2007st}.

In particular, in the limit $\Delta\tau\ll |\tau_0|$ the most important difference between (\ref{eqn:wact}) and the exact action~(\ref{eqn:vact}) comes from the term proportional to $z'/z$. Therefore the analysis of the action~(\ref{eqn:sigact}) will be consistent with the exact one of section~\ref{gauge} as long as $z'/z\sim \Delta c_s/\Delta\tau\lesssim \cH$ (where we have used $c_s\approx 1$), {\em i.e.}, 
\be\label{eqn:2cnstrnt}
|\Delta c_s|\lesssim \frac{\Delta\tau}{|\tau_0|} \,.
\ee
We expect that the regime of validity of the decoupling limit at the quadratic level will correspond to that at the cubic level. In other words, if the cubic Lagrangian in the decoupling limit turns out to be non vanishing, then it will give the dominant contribution to the three-point function up to ${\cal O}(\sqrt{\epsilon})$ corrections. This procedure has been shown to give correct results, {\em e.g.}, in the regime of small speed of sound~\cite{Cheung:2007st} and has been invoked for the study of models with resonant nongaussianities~\cite{Leblond:2010yq}. While a formal proof of the validity of the decoupling approximation for the cubic Lagrangian is beyond the scope of the present paper, we believe that the explicit check we have provided for the quadratic Lagrangian, together with the examples described above, supports the use of this approach.

Let us then move on to the cubic action:
\bea
S_\sigma^{(3)}&=&\int\d\tau\,\d^3 x\,\frac{M_p^2}{a}\,\Big\{\cH^2\mu'\sigma\sigma'{}^2+\cH^2\mu\,\sigma'{}^3-\cH^2\mu\,\sigma'(\pt_i\sigma)^2 \nn\\
&&\hspace{70pt} + {\cal O}(\epsilon\,\cH^2)\Big\} \equiv \int\d\tau\,(-H_{\rm int}) \,,
\eea
and the bispectrum is given by
\begin{align}\label{eqn:bispgf}
&\langle{\cal R}(\vec k_1,\tau)\,{\cal R}(\vec k_2,\tau)\,{\cal R}(\vec k_3,\tau)\rangle\\
&=-i\,H^3\int^\tau\d\tau'\langle\left[\sigma(\vec k_1,\tau)\,\sigma(\vec k_2,\tau)\,\sigma(\vec k_3,\tau),\,H_{\rm int}(\tau')\right]\rangle\,, \nn
\end{align}
where we have used ${\cal R}=-H\,\sigma$.

Before we go ahead and compute (\ref{eqn:bispgf}) we must make sure that we trust our perturbative expansion. The most relevant term in the cubic action is the term proportional to $\mu'$, let us denote it by $S^{(3)}_{\mu'}$. Perturbation theory will be valid if $S^{(3)}_{\mu'}\ll S^{(2)}_\sigma$, with
\begin{equation}
\frac{S^{(3)}_{\mu'}}{S^{(2)}_\sigma}
\sim \frac{\mu'}{\left(\epsilon+\mu\right)}\,\frac{\sigma}{a}
\sim \frac{\Delta\mu}{\Delta\tau}\,\frac{c_s^2}{\epsilon}\,\frac{H\sigma}{{\cal H}} 
\simeq 10^{-5}\,|\Delta c_s|\,\frac{|\tau_0|}{\Delta\tau}\,,
\end{equation}
where in the last equality we have used the fact that $c_s\simeq 1$, and where $H\sigma\simeq 10^{-5}$ is the amplitude of the perturbations. Let us also note that $\Delta\tau/|\tau_0|$ is the duration of the process in units of Hubble time. We thus find that as long as $\frac{\Delta\tau}{|\tau_0|}\gg 10^{-5}\,|\Delta c_s|$ we trust our perturbative expansion. Since this condition is less stringent than (\ref{eqn:2cnstrnt}), (\ref{eqn:2cnstrnt}) is the only condition we require for the validity of our analysis.

%%%%
\subsection{Equilateral case}
%%%

\begin{figure}
  \begin{center}
    \resizebox{.5\textwidth}{!}{\includegraphics{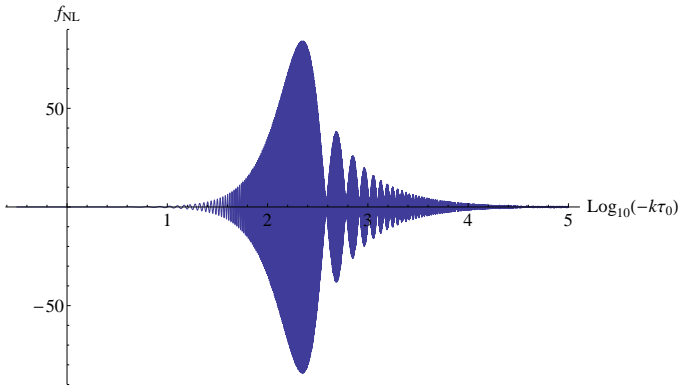}}
  \end{center}
  \caption{Equilateral $\fnl$ for $\Delta c_s=-.005$ and $\Delta\tau=-.01\,\tau_0$.}
  \label{bs}
\end{figure}

We are now in position to evaluate the three-point function. We first consider the equilateral configuration. With $\sigma_i=\frac{\hat v_i}{Hz}$, we have
%%%%
\begin{widetext}
\bea\label{eqn:equilfnl}
\fnl^{\rm eq}(k)&=&\frac{5\,M_p^2H^4}{6\,\pi^4}\frac{k^6}{{\cal P}(k)^2}\,\epsilon\,\Delta c_s\Big[-i\, \Big\{\int_{\tau_0}^{\tau_0+\Delta\tau}\frac{\d\tau}{\tau}\Big(f(\sigma_2'{}^*)^3 - \frac{k^2}{2}f\sigma_2'{}^*(\sigma_2^*)^2 + f'\sigma_2^*(\sigma_2'{}^*)^2\Big) \nn\\
&&\hspace{120pt}+ \int_{\tau_0+\Delta\tau}^0\frac{\d\tau}{\tau}\Big((\sigma_3'{}^*)^3 - \frac{k^2}{2}\sigma_3'{}^*(\sigma_3^*)^2\Big)\Big\}\cdot\sigma_3(\tau\to0)^3 + {\rm c.c.}\,\Big] \nn\\
&=&\frac{50}{81}\,\Delta c_s\,\Big[\frac{21}{20}+\frac{2}{\Delta\tau^5}\Big\{\Big(\frac{\tau_0\,\Delta\tau\,(3\,\tau_0-4\,\Delta\tau)}{k^2}+\frac{8\,\tau_0-7\,\Delta\tau}{k^4}\Big)\cos 3k\tau_0 \nn\\
&&\hspace{50pt}+\Big(\frac{\Delta\tau\,(\tau_0+\Delta\tau)\,(3\,\tau_0+7\,\Delta\tau)}{k^2}-\frac{8\,\tau_0+15\,\Delta\tau}{k^4}\Big)\cos 3k(\tau_0+\Delta\tau) \nn\\
&&\hspace{50pt}-\Big(\frac{3\,\tau_0^2\,\Delta\tau^2}{2\,k}-\frac{12\,\tau_0^2-60\,\tau_0\,\Delta\tau+11\,\Delta\tau^2}{6\,k^3}+\frac{22}{3\,k^5}\Big)\sin 3k\tau_0 \\
&&\hspace{50pt}+\Big(\frac{3\,\Delta\tau^2\,(\tau_0+\Delta\tau)^2}{2\,k}-\frac{12\,\tau_0^2+84\,\tau_0\,\Delta\tau+83\,\Delta\tau^2}{6\,k^3}+\frac{22}{3\,k^5}\Big)\sin 3k(\tau_0+\Delta\tau) \Big\}\Big]\,.\nn
\eea
\end{widetext}
%%%%
It can easily be found that the maximum of the feature is
\be\label{eqn:scalfnl}
f_{\rm N\hspace{-.5pt}L,max}^{\rm eq}=1.7\,|\Delta c_s|\,\frac{\tau_0^2}{\Delta\tau^2}\,\Big\{1+{\cal O}\Big(\frac{\Delta\tau}{\tau_0}\Big)\Big\}\,,
\ee
and occurs at 
\be\label{eqn:peakk}
k_{\rm max}=\frac{2.2}{\Delta\tau}\,\Big\{1+{\cal O}\Big(\frac{\Delta\tau}{\tau_0}\Big)\Big\}\,.
\ee
A plot of $\fnl^{\rm eq}$ for $\Delta c_s=-.005$ and $\Delta\tau=-.01\,\tau_0$ is shown in Fig.~\ref{bs}. The beats appearing in Fig.~\ref{bs} originate from the fact that the function $\mu(\tau)$ is not infinitely differentiable. We have checked numerically that the use of a smooth $\mu(\tau)\propto \left(\tanh\left(\frac{\tau-\tau_0}{\Delta\tau}\right)+1\right)$ leads to the disappearance of the beats without modifying the qualitative behavior of the three-point function. In particular, the scalings~(\ref{eqn:scalfnl}) and~(\ref{eqn:peakk}) found above are unaffected (up to an overall ${\cal {O}}(1)$ factor) by the choice of a ${\cal C}^2$-function $\mu(\tau)$.

Note that the bispectrum~(\ref{eqn:equilfnl}) does not vanish in the limit $k\to\infty$, that is, $c_s\neq 1$ leads to nonvanishing nongaussianity even in the decoupling limit. The limiting value agrees with that found in \cite{Seery:2005wm,Chen:2006nt} for constant $c_s$.\\

%%%
\subsection{Squeezed limit}
%%%

In the squeezed limit($k_1=k_2=k_L \gg k_3=k_S$), the contribution to $\fnl$ from the $\mu'\,\sigma\,\sigma'{}^2$ interaction is
%
%%%%
\begin{widetext}
\begin{align}\label{eqn:sqfnl}
\fnl^{\rm sq}(k_L,k_S)=&-\frac{75}{4}\,\frac{\Delta c_s}{\Delta\tau^5}\,\Big\{2\Big(\frac{\tau_0\Delta\tau^2}{3\,k_L^2}-\frac{\tau_0-2\,\Delta\tau}{k_L^4}\Big)\cos 2k_L\tau_0 -\Big(\frac{\Delta\tau^2(\tau_0+\Delta\tau)}{3\,k_L^2}-\frac{\tau_0+3\Delta\tau}{k_L^4}\Big)\cos 2k_L(\tau_0+\Delta\tau) \nn\\
&+\Big(\frac{\Delta\tau\,(2\,\tau_0-\Delta\tau)}{2\,k_L^3}+\frac{5}{2\,k_L^5}\Big)\sin 2k_L\tau_0+\Big(\frac{\Delta\tau\,(2\tau_0+3\Delta\tau)}{2\,k_L^3}-\frac{5}{2\,k_L^5}\Big)\sin 2k_L(\tau_0+\Delta\tau) \Big\} + {\cal O}\Big(\frac{k_S}{k_L}\Big)\,.
\end{align}
\end{widetext}
%%%%
Its maximum,
\be\label{eqn:sqfnlm}
f_{\rm N\hspace{-.5pt}L,max}^{\rm sq}=1.3\,|\Delta c_s|\,\frac{|\tau_0|}{\Delta\tau}\,\left\{1+{\cal O}\left(\frac{\Delta\tau}{\tau_0}\right)\right\}\,,
\ee
occurs at 
\be
k_{\rm max}=\frac{2.5}{\Delta\tau}\,\left\{1+{\cal O}\left(\frac{\Delta\tau}{\tau_0}\right)\right\}\,.
\ee
Note that it is smaller than that of the equilateral case -- as expected, since the amplification of the $\sigma$ modes occurs deep into the horizon -- by a factor of $\Delta\tau/|\tau_0|$. The leading contributions to $\fnl$ from the other interactions are suppressed by ${\cal O}\left(k_S/k_L\right)$, and therefore negligible. Using (\ref{eqn:sqfnl}) and (\ref{eqn:sd}), one can also check that the consistency relation of \cite{Creminelli:2004yq}
\be
\fnl^{\rm sq}(k_L,k_S)=-\frac{5}{12}\frac{\d\ln{\cal P}(k)}{\d\ln k}\Big|_{k=k_L}\,,
\ee
holds up to ${\cal O}\Big(\Delta c_s^2,\frac{k_S}{k_L}\Big)$ corrections.

%%%%%%%%%%%%%%%
%%%%%%%%%%%%%%%
%%%%%%%%%%%%%%%
%%%%%%%%%%%%%%%

%%%
\subsection{General case}
%%%

Figure~\ref{bs} shows that the behavior of $\fnl^{\rm eq}(k)$ is given by high frequency oscillations modulated by an envelope function. We have checked that the envelope of $\fnl^{\rm eq}$ can be written as
\be\label{eqn:eqlenv}
A_{\rm eq} + \sqrt{B_{\rm eq} + C_{\rm eq} \sin (3k\Delta\tau+\beta_{\rm eq})}\,,
\ee
with $A_{\rm eq}$, $B_{\rm eq}$, $C_{\rm eq}$ and $\beta_{\rm eq}$ functions of $k$, $\tau_0$ and $\Delta\tau$. Since we factored out the high frequency parts, $e^{3k\tau_0}$, to obtain (\ref{eqn:eqlenv}), $\fnl^{\rm eq}$ may be written as
\be
\fnl^{\rm eq} = A_{\rm eq} + \sin(3k\tau_0 + \alpha_{\rm eq}) \sqrt{B_{\rm eq} + C_{\rm eq} \sin (3k\Delta\tau+\beta_{\rm eq})}\,.
\ee
In the squeezed limit, we get the similar result with $3\,k$ replaced by $2k_L$. 

The above results can be generalized to arbitrary configurations of $k_1$, $k_2$ and $k_3$. The complete expression for $\fnl(k_1,k_2,k_3)$ is too lengthy to be explicitly presented, and it can be schematically written as
\be\label{eqn:gfnl}
\fnl(k_1,k_2,k_3) = A + \sin(K\tau_0 + \alpha) \sqrt{B + C \sin (K\Delta\tau+\beta)}\,,
\ee
where $K=k_1+k_2+k_3$, and $A$, $B$, $C$, $\alpha$ and $\beta$ are slowly varying functions of $k_1$, $k_2$, $k_3$, $\tau_0$ and $\Delta\tau$. This might be useful as a template for data analysis. We also checked that the $\Delta\tau\to0$ limit of (\ref{eqn:gfnl}) took the form of
\be
\fnl(k_1,k_2,k_3) \;{\underset{\Delta\tau\to0}{\to}}\; \tilde A + \tilde B \sin (K\tau_0+\tilde\beta) \,,
\ee
which agrees with the Ansatz proposed by, \eg, \cite{Chen:2008wn}.

%%%%%%%%%%%%%%%%%%%%%%%%%%%%%%%%%
\section{Discussion}\label{disc}%
%%%%%%%%%%%%%%%%%%%%%%%%%%%%%%%%%

We have shown that a sudden change in the speed of sound generates various distinct features in the primordial spectrum of scalar perturbations. 

First, due to the fact that the spectrum of perturbations is inversely proportional to the speed of sound, the normalization of the two-point function for modes that were deep inside the horizon before the transition took place is offset by an amount $\simeq 1-\Delta c_s$ with respect to that of the modes that were already far outside the horizon at the same instant.

Second, the two-point function of the scalar perturbations shows an oscillating pattern of amplitude $\Delta c_s$ that dies off at large $k$. But oscillations last all the way to $k\to \infty$ for an infinitely sharp jump -- in agreement with the results of~\cite{Bean:2008na,Nakashima:2010sa}.

Third, the three-point function (in the equilateral configuration) shows a feature of maximum amplitude $\sim \Delta c_s\,\left(\tau_0/\Delta\tau\right)^2$ at scales $k\sim\Delta\tau^{-1}$. In particular, it is quadratically divergent at large $k$ in the case of an instantaneous transition ($\Delta\tau=0$). An analogous behavior has been recently found in~\cite{Adshead:2011jq}, who have studied a rapid transition in the inflationary potential and have shown that the seemingly violent UV divergence in momentum space corresponds, due to its oscillatory nature, to a much milder logarithmic divergence in real space.

The three-point correlation function found in the present paper shares many properties with that found in~\cite{Adshead:2011jq}. As a consequence, the discussion of the detection prospects for such a three-point function  given in~\cite{Adshead:2011jq} applies to our scenario as well: if the scale $1/|\tau_0|$ roughly corresponds to current size of the horizon and the duration $\Delta\tau$ of the transition is short enough, a feature like (\ref{eqn:equilfnl}) or (\ref{eqn:sqfnl}) in the CMB bispectrum might be detectable in the upcoming surveys.

Let us conclude by emphasizing that the parameter controlling the speed of sound is just {\em one} of many parameters appearing in the effective field theoretical description of the inflationary perturbations~\cite{Cheung:2007st}. It would be interesting to  investigate whether a sudden change in some other parameters ({\em e.g.}, those controlling the behavior of the tensor modes) would result in novel observable features in the CMB or in the primordial spectrum of gravitational waves.

%%%
\acknowledgements We thank Guido D'Amico for useful discussion. This work is partially supported by the U.S. National Science Foundation grant PHY-0555304.
%%%

\end{document}